\documentclass[aps,prb,reprint,showkeys]{revtex4-1}

\usepackage{graphicx}
\begin{document}

\title{On the nature of change in Ni oxidation state in BaTiO$_3$--SrTiO$_3$ system}


\author{A. I. Lebedev}
\email[]{swan@scon155.phys.msu.ru}
\author{I. A. Sluchinskaya}
\affiliation{Physics Department, Moscow State University, Moscow, 119991, Russia}

\date{\today}

\begin{abstract}
XAFS studies of Ni-doped Ba$_{1-x}$Sr$_x$TiO$_3$ solid solution reveal that the Ni
oxidation state changes from 4 in SrTiO$_3$ to 2.5 in BaTiO$_3$ when varying $x$.
This change is accompanied by a noticeable change in the interatomic Ni--O distances
in the first shell. The first-principles calculations show that nickel creates an
impurity band in the forbidden band gap of BaTiO$_3$ and SrTiO$_3$, which explains
the appearance of intense absorption of Ni-doped samples in the visible region.
The analysis of the electronic structure of doped crystals and calculations of
the oxygen vacancy formation energy in them show that different oxidation states
of Ni in SrTiO$_3$ and BaTiO$_3$ can be explained by different formation energies
of the oxygen vacancies in these compounds.

\texttt{DOI: 10.1080/00150193.2016.1198196}
\end{abstract}


\maketitle

\section{Introduction}

Ferroelectric oxides with the perovskite structure are widely used in modern
electronics for production of multilayer ceramic capacitors, various
piezoelectric devices, and pyroelectric infrared
detectors.~\cite{Uchino2009,IntJApplCeramTechnol.2.1,Ferroelectrics.33.193}
Thin films of these materials are currently of great technological interest
for tunable capacitors~\cite{JElectroceram.11.5} and dynamic random access
memory (DRAM).~\cite{Uchino2009,AnnuRevMaterSci.28.79}

Barium titanate BaTiO$_3$, which is one of the end-member compound in the
BaTiO$_3$--SrTiO$_3$ system, undergoes successive ferroelectric phase
transitions from cubic $Pm3m$ to tetragonal $P4mm$ and further to
orthorhombic $Amm2$ and rhombohedral $R3m$ phases with decreasing temperature.
At 1460$^\circ$C it undergoes a structural phase transition to the hexagonal
$P6_3/mmc$ phase; the temperature of this transition can be controlled by
doping BaTiO$_3$ with vacancies and various impurities.~\cite{ProcPhysSoc.76.763}
The other end-member compound, strontium titanate SrTiO$_3$, is an incipient
ferroelectric, in which the softening of the TO mode is observed with
decreasing temperature, but the samples remain paraelectric up to the lowest
temperatures.~\cite{PhysRevB.19.3593}  At $\sim$105~K SrTiO$_3$ undergoes an
antiferrodistortive phase transition into the $I4/mcm$ phase. Barium titanate
and strontium titanate form a continuous series of solid solutions
Ba$_{1-x}$Sr$_x$TiO$_3$. When increasing the amount of SrTiO$_3$, the temperatures
of all ferroelectric phase transitions are decreased and near $x = 0.8$ they
merge into one phase transition.~\cite{PhysRevB.54.3151} These solid solutions
have been extensively studied because of their potential applications in various
microelectronic devices.

Recently, perovskite ferroelectrics have attracted increasing attention
because of their possible application in a new type of solar energy converters
based on the bulk photovoltaic effect. Since oxide perovskites have a relatively
large band gap ($\sim$3~eV), to match their absorption spectra to the solar
radiation spectrum they can be doped with $3d$-elements, which create so-called
color centers.~\cite{JInorgNuclChem.43.1499} Theoretical
studies~\cite{PhysRevB.83.205115} have shown that the doping of a relative
compound, PbTiO$_3$, by vacancy-compensated divalent impurities with the $d^8$
electron configuration (Ni, Pd, Pt at the $B$~site) can reduce the band gap to
the levels optimal for effective energy conversion.

Strontium titanate doped with $3d$-impurities (Mn, Co, Fe, and Ni) has recently
been studied using XAFS
spectroscopy.~\cite{JETPLett.89.457,JAdvDielectrics.3.1350031,PhysSolidState.56.449}
These studies have shown
that Ni is one of the most promising dopants from the viewpoint of the effective
sunlight absorption. The samples of Ni-doped strontium titanate were nearly
black, the Ni ions were shown to substitute for Ti and, which was the most
unusual, they were tetravalent. Unfortunately, SrTiO$_3$ is an incipient
ferroelectric and its doping with nickel does not result in the appearance
of ferroelectricity. This is why it is more interesting to study Ni-doped BaTiO$_3$.

Nickel is an important material for making electrical contacts to ferroelectric
ceramics, and this is why the initial attention was paid to such questions as
the Ni solubility in BaTiO$_3$ and its influence on the Curie temperature,
dielectric constant, and other properties of BaTiO$_3$ as well as to the
influence of nickel on the transition to the nonpolar $P6_3/mmc$ phase. It
was shown that the incorporation of Ni decreases the transition temperature
into the hexagonal phase, however, the solubility of nickel and its
concentration needed for this transition strongly depend on the synthesis
conditions, and the available data are very
contradictory.~\cite{ProcPhysSoc.76.763,MaterChemPhys.105.320,
JAlloysComp.481.559,SolidStateCommun.191.19,JPhysCondensMatter.23.115903}
It was found that Ni in BaTiO$_3$ acts as an acceptor,~\cite{SolidStateIonics.73.139}
and the doping of barium titanate with nickel results in a decreasing of both the
dielectric constant and the Curie temperature as well as in a smearing of the
ferroelectric phase transition in BaTiO$_3$ with increasing Ni
concentration.~\cite{JElectroceram.18.183,JElectroceram.21.394,IndJEngMaterSci.16.390}
The transformation kinetics between the cubic and hexagonal phases in BaTiO$_3$
doped with different impurities was studied in Ref.~\onlinecite{ProcPhysSoc.76.763}.
It was shown that nickel promotes the transition into the hexagonal phase,
whereas the doping of BaTiO$_3$ with strontium, in contrast, prevents this
transition. In Ref.~\onlinecite{MaterChemPhys.105.320} it was noticed that
doping of BaTiO$_3$ with nickel changes its color to dark brown.

The information about the oxidation state and structural position of Ni in
BaTiO$_3$ has been obtained mainly from EPR studies and is also very
controversial. The lines observed in EPR spectra have been attributed to
the Ni$^+$ ions at the $B$~sites,~\cite{JPhysCondensMatter.19.496214}
off-center Ni$^+$ ions at the $A$~sites.~\cite{SolidStateCommun.116.133} In
Ref.~\onlinecite{JPhysCondensMatter.23.115903}, EPR studies of Ni-doped hexagonal
BaTiO$_3$ have found Ni$^{3+}$ ions that replaces Ti$^{4+}$ in two different
positions (Ti(1) and Ti(2)); however, these centers can be associated with a
maximum of 5\% of the nominal amount of nickel, i.e. most of the Ni ions in
the sample are in an EPR inactive state. The literature data on the oxidation
state and structural position of nickel in SrTiO$_3$ have been summarized
in our papers;~\cite{JAdvDielectrics.3.1350031,PhysSolidState.56.449}  no data
on the subject for the Ba$_{1-x}$Sr$_x$TiO$_3$ solid solution was found.

One can see that, on the one hand, the optical properties of Ni-doped BaTiO$_3$
showing the strong absorption in the visible region in combination with its
ferroelectric properties suggest that this material can be used in solar
energy converters based on the bulk photovoltaic effect. On the other hand,
despite the fact that BaTiO$_3$(Ni) has long been studied, the literature
data on the Ni solubility, critical Ni concentration needed to transform
BaTiO$_3$ into the hexagonal phase, its oxidation state, and structural
position of Ni are very contradictory. Since the electronic transitions in
Ni-related color centers are determined by the nickel oxidation state and
the microscopic structure of these centers, in this work we used XAFS
spectroscopy to determine the properties of the Ni impurity in the
SrTiO$_3$--BaTiO$_3$ system. The obtained data will be used to propose an
adequate model for first-principles calculations, which will be used later
to explain the observed optical properties of Ni-doped Ba$_{1-x}$Sr$_x$TiO$_3$
and to optimize the preparation conditions for obtaining material suitable
for efficient solar energy converters.

\section{Samples, experimental and calculation techniques}

Samples of SrTiO$_3$, Ba$_{0.8}$Sr$_{0.2}$TiO$_3$, and BaTiO$_3$ doped with
0.5--3\% Ni were prepared by the solid-state reaction method at 1500$^\circ$C.
The starting components were BaCO$_3$, SrCO$_3$, nanocrystalline TiO$_2$
obtained by hydrolysis of tetrapropylorthotitanate and dried at 500$^\circ$C,
and Ni(CH$_3$COO)$_2$$\cdot$4H$_2$O. The components were weighed in required
proportions, grinded in acetone, and annealed in air at 1100$^\circ$C in
alumina crucibles for 4--8 h. The resulting powder was grinded again and
re-annealed in air at 1500$^\circ$C for 2 h. In order to incorporate the
impurity into the $B$ site, the composition of the samples was intentionally
deviated from stoichiometry toward the Ba excess. The phase composition of
the samples was checked by X-ray method. The samples were single-phase and
had cubic or hexagonal perovskite structures at 300~K. The synthesis of the
NiTiO$_3$ and BaNiO$_{3-\delta}$ reference compounds was described in
Ref.~\onlinecite{PhysSolidState.56.449}.

Measurements of X-ray absorption spectra in regions of the extended fine structure
(EXAFS) and near-edge structure (XANES) were carried out at the KMC-2 station of
the BESSY synchrotron radiation source at the Ni $K$-edge (8.34~keV) in
fluorescence mode at 300~K. The incident radiation was monochromatized using
two-crystal Si$_{1-x}$Ge$_x$(111) monochromator. The intensity of the radiation
incident on the sample was measured using an ionization chamber, the intensity
of the excited fluorescence was measured with a R\"ONTEC X-flash
energy-dispersive silicon detector with a working area of 10~mm$^2$.

EXAFS spectra were processed with the widely used \texttt{IFEFFIT} software
package.~\cite{IFEFFIT} The EXAFS function was extracted from the experimental
spectra
using the \texttt{ATHENA} program, and its fitting to the theoretical curve
calculated for a given structural model was performed using the \texttt{ARTEMIS}
program. The amplitude and phase shifts for all paths of single and multiple
scattering were calculated using the \texttt{FEFF6} code.

Modeling of the geometry and electronic structure of Ni-doped BaTiO$_3$ and
SrTiO$_3$ was performed within the first-principles density functional theory
using the \texttt{ABINIT} software. The calculations were carried out on the
40-atom (SC) and 80-atom (FCC) supercells, in which one of Ti atoms was replaced
by the Ni atom (the Ni concentration was 12.5 and 6.25\%, respectively). The
symmetry of supercells was cubic for both SrTiO$_3$ and BaTiO$_3$. Since nickel
has a partially filled $d$-shell, the PAW pseudopotentials~\cite{ComputMaterSci.81.446}
and LDA+$U$ approximation~\cite{JPhysCondensMatter.9.767} were used in the
calculations. The $U$ and $J$ parameters
describing the Coulomb and exchange interaction within the $d$-shell were
taken from the literature as $U = 5$~eV, $J = 0.9$~eV equal to typical values
of these parameters for Ni; the changes by 20\% of these parameters were
shown to have a small influence on the results. The cut-off energy was 30~Ha,
the integration over the Brillouin zone was carried out on a 4$\times$4$\times$4
Monkhorst-Pack mesh. Relaxation of the lattice parameters and atomic positions
in the supercells was stopped when the Hellmann-Feynman forces were below
1$\times$10$^{-5}$~Ha/Bohr.

The modeling on the 40-atom and 80-atom supercells without oxygen vacancies
gave the results for the Ni$^{4+}$ oxidation state. In order to change the
oxidation state to Ni$^{2+}$, we used a trick~\cite{Ferroelectrics.206.69}
where two extra electrons
were added to the system to change the filling of the $d$-shell. Although the
system is not electrically neutral in this case, the tests have shown that the
resulting density of states and the interatomic distances were very close to
those calculated for the model in which the Ni$^{2+}$ oxidation state was
obtained by the addition of an oxygen vacancy at a distance of 5.8~{\AA} from
the Ni atom.

\section{Experimental results}

\begin{figure}
\includegraphics{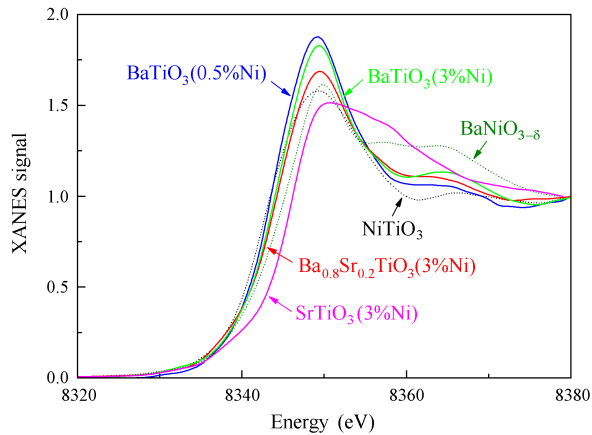}
\caption{\label{fig1}(Color online) XANES spectra of four samples in the
BaTiO$_3$--SrTiO$_3$ system and two reference compounds.}
\end{figure}

To determine the Ni oxidation state, the position of the absorption edge in
XANES spectra of the samples was compared with those of the reference compounds.
XANES spectra of SrTiO$_3$, Ba$_{0.8}$Sr$_{0.2}$TiO$_3$, and BaTiO$_3$ samples
doped with nickel and two reference compounds (BaNiO$_{3-\delta}$ and NiTiO$_3$)
are shown in Fig.~\ref{fig1}. It is seen that in BaTiO$_3$(0.5\%Ni) sample, the
absorption edge is closest to the absorption edge of the NiTiO$_3$ reference compound
in which the oxidation state of nickel is 2+. In hexagonal BaTiO$_3$(3\%Ni) and
cubic Ba$_{0.8}$Sr$_{0.2}$TiO$_3$(3\%Ni) samples, the positions of the absorption
edge are almost the same and are shifted by 0.7~eV with respect to the absorption
edge of NiTiO$_3$ toward the absorption edge of the BaNiO$_{3-\delta}$ reference
compound, in which the oxidation state of nickel is 3.4+.%
    \footnote{The Ni oxidation state in this sample was determined in
    Ref.~\onlinecite{PhysSolidState.56.449}.}
The absorption edge in Ni-doped SrTiO$_3$ sample is by 2.8~eV above the NiTiO$_3$
edge. If we assume that the Ni oxidation state in SrTiO$_3$ is 4+ (see the
discussion in Ref.~\onlinecite{PhysSolidState.56.449}), we can conclude that
the average oxidation state of
Ni is $\sim$2.3 in BaTiO$_3$(0.5\%Ni) and $\sim$2.5 in both BaTiO$_3$(3\%Ni) and
Ba$_{0.8}$Sr$_{0.2}$TiO$_3$(3\%Ni). This means that the most of the nickel ions
in the latter samples are in 2+ oxidation state, and only a fraction of them are
in 3+ or 4+ states.

\begin{table}
\caption{\label{table1}Local structure of investigated samples doped with 3\% Ni.}
\begin{ruledtabular}
\begin{tabular}{cccc}
Shell & \multicolumn{3}{c}{$R_i$ (\AA)} \\
\cline{2-4}
      & SrTiO$_3$ & Ba$_{0.8}$Sr$_{0.2}$TiO$_3$ & BaTiO$_3$$^a$ \\
\hline
$R_{\rm Ni-O(I)}$ & 1.914$\pm$0.004 & 2.106$\pm$0.008 & 2.069 \\
$R_{\rm Ni-O(II)}$ & & 2.438$\pm$0.021 \\
$R_{\rm Ni-Ba}$ & 3.342$\pm$0.006 & 3.428$\pm$0.005 \\
$R_{\rm Ni-Ti}$ & 3.877$\pm$0.004 & 3.998$\pm$0.005 \\
\end{tabular}
\end{ruledtabular}
{\footnotesize $^a$Hexagonal structure. \hfill}
\end{table}

In order to determine the structural position of nickel in the samples, the EXAFS
spectra were additionally analyzed. The best agreement between the calculated
and experimental spectra for all samples was obtained in a model in which the Ni
impurity atoms substitute for Ti$^{4+}$ ions. In the case where Ni is in the 2+
oxidation state, the electrical neutrality of the sample is provided by a distant
oxygen vacancy $V_{\rm O}$. The interatomic distances in the first shell in
investigated samples are given in Table~1. It is seen that there is a significant
difference in Ni--O distances in the first shell for different Ni oxidation states
despite the fact that Ni atoms substitute for Ti atoms and are on-center in both
SrTiO$_3$ and BaTiO$_3$. The obtained Ni--O distances are close to the sum of
ionic radii of O$^{2-}$ and Ni ions in corresponding oxidation states, so that
the XANES and EXAFS data agree well.

\section{Results of first-principles calculations and discussion}

A surprising fact established in this work is a significant difference in the Ni
oxidation state in two related compounds: it is about 2+ in both BaTiO$_3$ and
Ba$_{0.8}$Sr$_{0.2}$TiO$_3$ and 4+ in SrTiO$_3$. This effect may result from the
difference in the $V_{\rm O}$ formation energies and from the difference in
relative positions of $V_{\rm O}$ donor levels and Ni acceptor levels in these
compounds. In order to clarify the nature of this effect, we calculated the
site-projected partial density of states (DOS) as well as the vacancy formation
energy in nickel-doped BaTiO$_3$ and SrTiO$_3$.

\begin{figure*}
\includegraphics{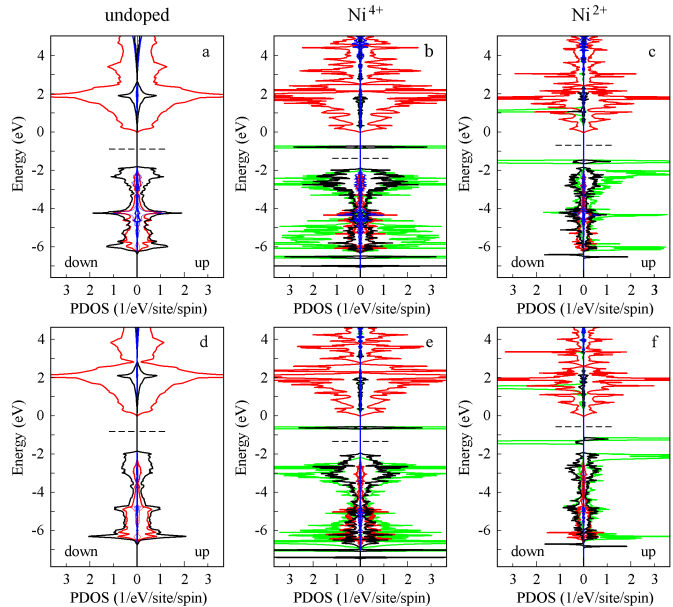}
\caption{\label{fig2}(Color online) Site-projected partial density of states for
undoped and Ni-doped samples of (a--c) BaTiO$_3$ and (d--f) SrTiO$_3$. Ti $3d$
states are shown by red, Ni $3d$ states are shown by green, O $2p$ states are
shown by black, and Ba $5d$ (Sr $4d$) states are shown by blue. The Fermi levels
are depicted by dashed lines.}
\end{figure*}

The results of the DOS calculations are presented in Fig.~\ref{fig2}. In the case
of tetravalent nickel, the ground-state structure is diamagnetic ($S = 0$) for
both BaTiO$_3$(Ni) and SrTiO$_3$(Ni). As follows from Fig.~\ref{fig2}, an impurity
band appears in the forbidden band gap of SrTiO$_3$ [Fig.~\ref{fig2}(e)] and
BaTiO$_3$ [Fig.~\ref{fig2}(b)]. In both compounds, the band is slightly shifted
toward the conduction band, but in SrTiO$_3$ it is located by 0.15--0.2~eV higher
than in BaTiO$_3$. The comparison of DOS of doped [Fig.~\ref{fig2}(b,e)] and undoped
[Fig.~\ref{fig2}(a,d)] compounds shows that the doping slightly increases the band
gap of the samples (by 82~meV for BaTiO$_3$ and by 91~meV for SrTiO$_3$ upon
adding of 6.25\% of Ni). It also increases the width of the valence band. The
Fermi level in doped crystals is located in the forbidden gap between the
impurity and the valence bands (dashed lines in the figure). The obtained results
disagree with the conclusion of Ref.~\onlinecite{PhysRevB.83.205115}  about
decreasing of the energy band gap upon Ni doping. The width of the impurity
band is $\sim$0.11~eV in SrTiO$_3$ and BaTiO$_3$ doped with 6.25\% Ni and
increases to 1.1--1.7~eV when doubling the impurity concentration.

For divalent nickel, the ground-state structure is paramagnetic ($S = 1$) in both
BaTiO$_3$(Ni) and SrTiO$_3$(Ni). Because of the appearance of magnetic moment,
the energy position of the Ni spin-up and spin-down $3d$ states is different:
the spin-up states are pushed down below the Fermi level, and the spin-down
states are shifted up and are located in the conduction band. In both compounds,
the impurity band is shifted toward the valence band, and the Fermi level is
located between the fully occupied spin-up states and the conduction band edge
[Fig.~\ref{fig2}(c) and \ref{fig2}(f)]. The energy splitting of the O $2p$ spin-up
and spin-down states indicates a partial magnetization of the oxygen ions located
in the vicinity of the paramagnetic Ni$^{2+}$ centers.

The formation energy of the oxygen vacancy in Ni-doped samples was calculated
as the difference between the total energy of the structure containing the Ni$^{4+}$
ion and the sum of the energy of the structure containing the Ni$^{2+}$--$V_{\rm O}$
complex and a half of the energy of the $O_2$ molecule in the triplet state.
It was 3.14~eV for SrTiO$_3$(Ni) and 2.69~eV for BaTiO$_3$(Ni).

We believe that the difference in the Ni behavior in SrTiO$_3$ and BaTiO$_3$ is
due to the difference in $V_{\rm O}$ formation energies in these materials.
Indeed, previous calculations of the energy levels of the oxygen
vacancies~\cite{IntJQuantChem.106.2173,ApplPhysLett.98.172901}
have shown that these levels are located in the upper half of the
forbidden gap and so are higher than the Ni acceptor levels. Therefore, the
filling of the Ni$^{4+}$ levels by electrons from the oxygen vacancies is
energetically favorable in both materials. On the other hand, our calculations
show that the formation energy of the oxygen vacancy in Ni-doped BaTiO$_3$ is
notably lower than in Ni-doped SrTiO$_3$. So, we can conclude that the observed
difference is associated with different formation energies of the oxygen vacancy
in these materials.

The proposed model can also qualitatively explain the effect of the impurity
concentration on the average oxidation state of Ni in doped samples. At low doping
level, the Ni impurity band is narrow, and its occupation by electrons is
determined by the relative energy positions of Ni and $V_{\rm O}$ levels. When
the impurity concentration increases, the energy bands formed from the Ni and
$V_{\rm O}$ levels become broad. Because of the overlapping of these bands,
the filling of the Ni impurity band asymptotically approaches that corresponding
to the 3+ oxidation state, in agreement with experiment.

It should be noted that our calculations of the electronic structure of a number
of nickel compounds have shown that the Ni--O interatomic distance is mainly
determined by the magnetic state of Ni rather than by its oxidation state. In
particular, it appeared that the Ni$^{2+}$--$V_{\rm O}$ complex with the
neighboring oxygen vacancy considered in Ref.~\onlinecite{PhysRevB.83.205115}
is diamagnetic and is
characterized by 4$\times$1.870+1$\times$2.144~{\AA} Ni--O distances, which are
very different from the distances obtained from the EXAFS data analysis. So,
only the paramagnetic Ni complexes with distant vacancies can explain the Ni--O
distance observed in experiment.

To check the proposed model, in future it would be interesting to study the
evolution of the Ni oxidation state upon change of the composition $x$ in the
whole BaTiO$_3$--SrTiO$_3$ system. The measurements of magnetic susceptibility
and determination of the spin concentration in the samples can be also useful
to check this model.

The location of the Ni impurity band near the middle of the forbidden gap of
SrTiO$_3$ and BaTiO$_3$ can explain the strong absorption observed in all
studied samples by electronic transitions between the impurity band and the
valence or conduction bands.

\section{Conclusions}

XAFS studies of Ni-doped Ba$_{1-x}$Sr$_x$TiO$_3$ solid solution have revealed
that the Ni oxidation state changes from 4 in SrTiO$_3$ to 2.5 in BaTiO$_3$
when varying $x$. This change is accompanied by a noticeable change in the
interatomic Ni--O distances in the first shell. The first-principles
calculations showed that nickel creates an impurity band in the forbidden band
gap of BaTiO$_3$ and SrTiO$_3$, which explains the appearance of intense
absorption of Ni-doped samples in the visible region. The analysis of the
electronic structure of doped crystals and calculations of the oxygen vacancy
formation energy in them showed that different oxidation states of Ni in
SrTiO$_3$ and BaTiO$_3$ can be explained by different formation energies
of the oxygen vacancies in these compounds.

\begin{acknowledgments}
This work was supported by Russian Foundation for Basic Research, grant No. 13-02-00724.
The authors would like to thank the BESSY staff for continuous support of our experiments.
\end{acknowledgments}

\providecommand{\BIBYu}{Yu}

\end{document}